\documentstyle[twoside,fleqn,espcrc2,draft]{article}

\newcommand{\AmS}{{\protect\the\textfont2
  A\kern-.1667em\lower.5ex\hbox{M}\kern-.125emS}}
\newcommand{\ssc}{\scriptscriptstyle}
\newcommand{\quattrova}{($\phi^{\scriptscriptstyle a},\,
c^{\scriptscriptstyle a},\,
\lambda_{\scriptscriptstyle  a},\,
{\bar c}_{\scriptscriptstyle a}$)}  %$
\newcommand{\be}{\begin{equation}}
\newcommand{\ee}{\end{equation}}
\newcommand{\bea}{\begin{eqnarray}}
\newcommand{\eea}{\end{eqnarray}}
\def \HT{{\widetilde{\mathcal H}}}
\def \LT{{\widetilde{\mathcal L}}}

\title{Quantization and Time}
\author{A.~A.~Abrikosov, Jr.\address{ITEP, Bol.~Cheremushkinskaya, 25;
        Moscow, 117259 Russia.}
and
Ennio Gozzi\address{Department of Theoretical Physics,
University of Trieste \\
Strada Costiera 11, Miramare-Grignano 34014, Trieste\\
and INFN, Trieste, Italy.}}

\begin{document}

\begin{abstract}
Starting from a functional formulation of classical mechanics, we show
how to perform its quantization by freezing to zero two Grassmannian
partners of time.
\end{abstract}

\maketitle

\section{INTRODUCTION}
This conference is centered on  {\it quantum gravity\/}
which seems to be one of the most difficult and unsolved problems
that physicists have faced in this century. The two components of the
problem are {\it quantum mechanics\/} and {\it gravity.\/}
Physicists have worked mostly on the second one that is gravity.
They have tried to modify gravity by going to supergravity, strings
and M-theory {\em etc.\/}  with the hope of obtaining a theory that, after
quantization, would be free of the ultraviolet problems of standard
gravity. We wonder why they have not thought more about the first
component of the problem, namely about quantum mechanics (QM). By "thought
more" we do not mean  problems like the measurement issue
{\em etc.\/}, but we mean the {\it geometrical aspects\/} of quantum
mechanics.  After all, before marrying quantum mechanics to gravity
which is the queen of geometrical theories, one should better
understand quantum mechanics from a more geometrical point of view. We mean
the following. Quantum mechanics is usually formulated via Hilbert space
kind of tools and space-time  apppears like a secondary concept. This, we think,
creates hidden conceptual difficulties in applying quantum mechanics
to space-time theories such as gravity. The standard procedure itself of
quantization seems to have nothing to do with space-time.
In this paper we shall not
change this picture. In particular  we shall {\it not\/}  
suggest things like quantization of time as one could have wrongly
guessed from the title of the talk. What we will show is that,
by formulating classical mechanics (CM) in a more modern way,\cite{enniocl},
the standard quantization rules become equivalent to freezing to zero
some Grassmannian partners of time. This is how "space-time" concepts enter 
the picture of quantization in our approach.

At the beginning of a new century it seems quite timely to start
rethinking about quantum mechanics. Remember how the last century started: on
{\it Dec.14th, 1900} M.~Planck presented the paper, \cite{planc}, which
contained the idea of the Planck constant and that day was called by
A.~Sommerfeld the "{\it birthday of quantum mechanics}", \cite{som}.

\section{FUNCTIONAL APPROACH TO CLASSICAL MECHANICS}

The "modern" formulation of  classical mechanics (CM) mentioned in the
introduction is actually a  {\it functional\/} approach to the old
{\it operatorial\/} version of CM proposed by Koopmann and von
Neumann, \cite{koop}. These authors, instead of using the Hamiltonian
and the Poisson brackets for the classical evolution of a system, 
used the well-known Liouville operator,
${\hat L}\equiv{\partial H\over\partial p} {\partial\over\partial q}-
{\partial H\over\partial q}{\partial\over\partial p},$
and the associated commutators. One can
even generalize their formalism to higher forms and get the Lie derivative
of the Hamiltonian flow, \cite{marsd}.  It was shown in
\cite{enniocl} that the operatorial formalism mentioned above  could
have a functional or path integral counterpart even at the level of
pure classical mechanics. The path integral \cite{enniocl} basically
assigns weight one to classical paths and zero to others.

Let us denote by ${\cal M}$ our phase space with $2n$ phase-space coordinates
$\varphi^{a}=(q^{\ssc j},p^{\ssc j})$ (the index "$a$" spans both $q$'s and
$p$'s) and by $H(\varphi)$ the Hamiltonian of the system. Then the
classical probability $P\bigl(\varphi_{\ssc f},t_{\ssc f}\vert
\varphi_{\ssc i},t_{\ssc i})$ of going from the initial phase-space 
configuration
$\varphi_{\ssc i}$ at time $t_{\ssc i}$ to the final one
$\varphi_{\ssc f}$ at time $t_{\ssc f}$, is nothing but:
\be
\label{eq:uno}
P(\varphi_{\ssc f},t_{\ssc f}\vert\varphi_{\ssc i},t_{\ssc i}\bigr)
=\delta\bigl[\varphi_{\ssc f}-{\tilde\phi}_{cl}(t_{\ssc f};\,
\varphi_{\ssc i}, t_{\ssc i})\bigr],
\ee
where ${\tilde\phi}_{cl}(t;\, \varphi_{\ssc i}, t_{\ssc i})$ is the classical
trajectory  starting at the moment $t_{\ssc i}$ from the phase-space point
$\varphi_{\ssc i}$.
These trajectories are solutions of the Hamilton equations of motion:
$\varphi^{a}=\omega^{ab}{\partial H\over \partial\phi^{b}}$ with
$\omega^{ab}$ being the standard symplectic matrix.  Slicing the time
interval $t_{\ssc f}-t_{\ssc i}$ into n smaller ones and doing some
manipulations on the Dirac delta in the RHS of eq.(1), one can
rewrite the probability as a path integral (for more detail see
ref.\cite{enniocl}):
\be
\label{eq:due}
P\bigl(\varphi_{\ssc f},t_{\ssc f}
\vert\varphi_{\ssc i},t_{\ssc i}\bigr) =
\int{\cal D}{\mu}\, \exp i\int_{t_{\ssc i}}^{t_{\ssc f}}{\widetilde{\cal
L}}\, dt,
\ee
where
${\cal D}\mu\equiv {\cal D}^{\prime\prime} \varphi^{a} {\cal D}
\lambda_{a} {\cal D}c^{a} {\cal D}{\bar c}_{a}$.
The notation ${\cal D}^{\prime\prime}$ indicates that first and last
integrations over $\varphi$ are not done. The  geometrical meaning of 
the $6n$
auxiliary variables $\lambda_{a}$,\, $c^{a}, \,{\bar c}_{a}$ (where the
$c$ and ${\bar c}$ variables are Grassmannian) is explained in detail in
\cite{enniocl}. The Lagrangian ${\widetilde{\cal L}}$ in eq.(2) and the
associated Hamiltonian are:
\bea
\label{eq:tre}
{\widetilde{\cal L}}&=&\lambda_{a}[{\dot\varphi}^{a}-\omega^{ab}
\partial_{b}H]+
i{\bar c}_{a}[\delta^{a}_{b}\partial_{t}-\omega^{ac}\partial_{c}\partial_{b}H]
c^{b}\nonumber\\
{\HT}&=&\lambda_a\omega^{ab}\partial_bH+i\bar{c}_a\omega^{ac}
(\partial_c\partial_bH)c^{b}
\eea
The connection with the operatorial formalism, \cite{koop}, can be
established immediately  and has been thoroughly studied in
ref.\cite{enniocl}. There it was shown that the variables $\varphi^{a}$ and
$c^{a}$ can be realized as multiplicative operators while
$\lambda_{a}$ and ${\bar c}_{a}$ are derivative ones. So one
could define a basis $|\varphi,c>$ as
${\hat\varphi}|\phi,c>=\varphi|\varphi,c>$ and ${\hat c}|\varphi,c>=
c|\varphi,c>$ and the kernel of propagation between these states would
have the following path integral representation:
\bea
\label{eq:quattro}
<\varphi_{\ssc f},c_{\ssc f};\,
t_{\ssc f} |  \varphi_{\ssc i},c_{\ssc i};\, t_{\ssc i}> =
\hspace{60pt}\\ \hspace{20pt} =  \int^{(\varphi_{\ssc f},
c_{\ssc f})}_{(\varphi_{\ssc i},c_{\ssc i})}~{\cal D}\mu^{\prime}~
exp~i\int\LT dt, \hfill
\nonumber
\eea
where ${\cal D}\mu^{\prime}$  is the same measure as in eq.(2) but without
initial and final integrations over $c^{a}$.
The relation between the probability, eq.(2),
and the amplitude, eq.(4), is:
\bea
\label{eq:cinque}
P\bigl(\varphi_{\ssc f},t_{\ssc f} \vert \varphi_{\ssc i},
t_{\ssc i}\bigr) = \hspace{110pt} \\
K\int |<\varphi_{\ssc f},c_{\ssc f};\,
t_{\ssc f} \vert
\varphi_{\ssc i},c_{\ssc i};\, t_{\ssc i}>|^{2} \,
dc_{\ssc i}dc_{\ssc f};
\nonumber
\eea
where $K$ is an appropriate constant. The reader may be puzzled why
both the probability, eq.(2), and the amplitude, eq.(4), have the
same path integral representation. This happens because both of them
evolve via the Liouville operator that contains only first order
derivatives (differently than the second order Schr\"odinger operator).

Now we would like to address the
question of how we should quantize CM once it is formulated this way?
Some attempts to answer it have been done in \cite{ennioqm}. 
In the next section we shall put forward a new idea.

\section{QUANTIZATION AND GRASSMANNIAN PARTNERS OF TIME.}
The whole $8n$ variables \quattrova~can be assembled into
a single variable as follows.
First we introduce two Grassmannian partners, $\theta$ and ${\bar\theta}$,
of the standard time $t$. This converts the base space (which is
$t$) to a {\em superspace:\/}  $(t,\, \theta,\,{\bar\theta})$. 
We can then put together all variables
~\quattrova ~in a single "super"-field variable $\Phi^{a}$ defined as follows:
\bea
\label{eq:sei}
\Phi^{a}(t,\theta,{\bar\theta})\equiv \phi^{a}(t)&+&\theta c^{a}(t)+\\
&+&{\bar\theta}\omega^{ab}{\bar c}_{b}(t)
+i{\bar\theta}\theta\omega^{ab}\lambda_{b}(t). \nonumber
\eea
The superfield $\Phi^{a}$ is a very useful tool and, unexpectedly,
it brings to light some interplay between CM and QM. Let us for example
replace the variable $\varphi^{a}$ with the superfield $\Phi^{a}$
into the original Hamiltonian $H(\varphi^{a})$ of
our classical system. Expanding it in $\theta$ and ${\bar\theta}$ we get:
\bea
\label{eq:sette}
H(\Phi^{\ssc a})=H(\varphi^{\ssc a})&+&
\theta{\partial H\over\partial\varphi^{a}}c^{a}+\\
&+&{
\bar\theta}{\partial H\over\partial\varphi^{a}}\omega^{ab}{\bar c}_{b}+
i\theta{\bar\theta}\HT,
\nonumber
\eea
and doing the same with the old action associated with $H(\varphi)$, that
is $S=\int (p{\dot q}-H)dt$, we get:
\be
\label{eq:otto}
S[\Phi^{\ssc a}]=S[\varphi^{\ssc a}]+\theta {\cal T}+{\bar\theta}{\cal
V}+i \theta\, {\bar\theta}\, [{\widetilde {\cal S}}+(s.t.)].
\ee
Here ${\cal V}$ and ${\cal T}$
are functionals that are of no interest for the moment and the
$(s.t.)$ is a surface term: $(s.t.)={1\over 2}(\lambda_{a}\varphi^{a}
+i{\bar c}_{a}c^{a})\vert^{t_{\ssc f}}_{t_{\ssc i}}$.
The last term, ${\widetilde{\cal S}}=\int\LT dt$, is the action which
enters our CM path integral, eq.(2), while the first term, $S(\varphi)$, is
the action of the QM path integral. The occurrence, in
the same supermultiplet, of both the  QM and CM actions cannot be an accident 
and must have some intriguing geometrical meaning.

Let us first notice that the classical weight ${\widetilde{\cal S}}$
can be obtained from the action $S[\Phi]$:
\be
\label{eq:nove}
{\widetilde{\cal S}}=\int S[\Phi]id\theta d{\bar\theta}-(s.t.).
\ee
This implies that eq.(4) can be rewritten as:
\bea
\label{eq:dieci}
<\varphi_{\ssc f},c_{\ssc f};
t_{\ssc f}\vert\varphi_{\ssc i},c_{\ssc i};t_{\ssc i}>=  \hspace{57pt} \\
\hspace{20pt} \int{\cal D}\Phi\, \exp i\int S[\Phi]\, id\theta \,
d{\bar\theta}-(s.t.);
\nonumber
\eea
where we have formally written the measure ${\cal D}{\mu}^{\prime}$ of
eq.(4) in terms of superfields $\Phi$. We can get rid of the surface term
$(s.t.)$ in the RHS of eq.(10) by performing a proper Fourier transform of
both sides. This Fourier transform involves some of the variables
$\varphi^{a}$ and their "momenta" $\lambda_{a}$ ($\lambda_{a}$ are momenta
if one uses the $\LT$ as lagrangian.) The result is that the $(q,\, p)$-variables
are replaced by the $(q,\, \lambda_{p})$ and $(c^{q},\, {\bar c}_{p})$
take the place of $(c^{q},\, c^{p})$. The index "$q$" stands for the first
$n$ indices "$a$" in \quattrova,  and "$p$" stands for the second ones. 
After the Fourier transform we obtain:
\be
\label{eq:undici}
<\Phi^{q}_{\ssc f};t_{\ssc f}\vert\Phi^{q}_{\ssc i};t_{\ssc i}>
=\int{\cal D}\Phi~e^{i\int
S[\Phi]\, id\theta \, d{\bar\theta}},
\ee
where the LHS of (11) stands for
\be
\label{eq:dodici}
<q_{\ssc f},\lambda_{p,\ssc f},c^{q}_{\ssc f},
{\bar c}_{p,\ssc f};t_{\ssc f}\vert q_{\ssc i},
\lambda_{p,\ssc i},c^{q}_{\ssc i},
{\bar c}_{p,\ssc i}; t_{\ssc i}>
\ee
which is the Fourier transform of the LHS of eq. (10). Note that the
variables entering the bra and ket of (12) are exactly those making up the
superfield $\Phi^{q}$ in the LHS of (11). At this point we cannot avoid
noticing the formal analogy between the {\it classical} relation (11) and
the standard {\it quantum}  one which is:
\be
\label{eq:tredici}
<\varphi^{q}_{\ssc f};t_{\ssc f}\vert\varphi^{q}_{\ssc i};t_{\ssc i}>
=\int{\cal D}\varphi~e^{{i\over\hbar}S[\varphi]} .
\ee
One goes from  eq. (11) to (13) by letting
$\theta,{\bar\theta}$ go to zero:
$\lim_{\theta,\, {\bar\theta}\rightarrow 0}\Phi^{\ssc q}
 = \varphi^{\ssc q}=q$. 
The whole quantization procedure can then be formally represented as:
\bea
\label{eq:quattordici}
<\Phi^{q}_{\ssc  f}\vert\int{\cal D}\Phi~
&\exp&{i\int S[\Phi]\, id\theta\, d{\bar\theta}}
\vert\Phi^{q}_{\ssc i}>\nonumber\\
% &\Downarrow
\mathrm{quantization:}
&\left \Downarrow\rule[3pt]{0pt}{10pt} \right.
&~\theta,{\bar\theta}\rightarrow 0\\
<q_{\ssc f}\vert\int{\cal D}\varphi &\exp&{{i\over\hbar}S[\varphi]}\vert
q_{\ssc i}>. \nonumber
\eea
One may wonder how $\hbar$ appears. First let us notice from eqs.(8)
and (9) that the dimension of the volume element $d\theta\, d{\bar\theta}$ is
the inverse of  action. A correct procedure of sending $\theta,\,
{\bar\theta}$ to zero must respect the dimension. This may be done by
sticking  $-{i\over\hbar} \delta(\theta)\, \delta({\bar\theta})$ into the
$d\theta\, d{\bar\theta}$ integration in the exponent in the RHS of (11).
The procedure can be made more rigorous and precise,
\cite{go2000}, by introducing the generating functional $Z_{\ssc cl}[J]$
associated to the classical path-integral, (11), where $J$ is a
supercurrent coupled to the superfield $\Phi$. By performing a
proper  limit of $\theta,{\bar\theta} \rightarrow 0$, it is
possible to show that $Z_{\ssc cl}[J]$ goes into the quantum generating
functional $Z[j]$ where $j$ is the component of the
supercurrent $J$ coupled to the variable $\varphi$.

The procedure that we have outlined above belongs to the same
category as geometric quantization, \cite{wood1}. The reason to say
this is that geometric quantization brings us from the Lie derivative of
the Hamiltonian flow to the Schr\"odinger operator. We achieved the
same by our procedure: we basically managed to go from the weight
${\widetilde{\cal S}}$ of the classical path integral (and the associated
$\HT$ which is the classical Lie derivative, \cite{enniocl}) to the weight
$S(\varphi)$ that generates the Schr\"odinger operator.  All the complex
machinery of geometric quantization, \cite{wood1}, is basically {\it
encapsulated\/} in the $\theta,\, {\bar\theta}\rightarrow 0$ procedure. The
reader expert in geometric quantization will notice that the issue of
"polarization" was naturally incorporated via the surface terms, $(s.t.)$,
of eq.(10) which {\it magically} appeared thanks to the superfield formalism.  
The surface terms plus the $\theta,\, {\bar\theta}
\rightarrow 0$ procedure naturally bring us from $(q,\, p)$-von Neumann states
to $q$-Schr\"odinger states and simultaneously generate the correct path
integral weight.  The beauty of the procedure is that it generates the
correct states and the correct weight at once.  Other polarizations can be
obtained in a similar way, \cite{go2000}.

\section{CONCLUSIONS}
If quantization is equivalent to sending the Grassmannian partners of time
to zero, we should ask ourselves what this means from a physical point
of view. First we should understand the physical meaning of
the variables $\theta,\,{\bar\theta}$. In our formalism, \cite{enniocl},
all Grassmannian variables in the target-space are forms,
for example $c^{a} \sim d\varphi^{a}$. Almost the same  happens at
the base-space level,\cite{go2000}, where $\theta \sim \sqrt{dt}$. Thus the
projector,  ${\theta\,{\bar\theta}\over \hbar}$, that we are putting in to
pass from CM to QM is equivalent to $(\Delta t)\sim \hbar$.  
This is a well-known fact \cite{sak} which is basically the
{\it central feature}  QM.

The second point that we are exploring, \cite{go2000}, is the following. In
our formulation of CM, \cite{enniocl}, we have found several {\it universal}
symmetries relating the bosonic variables to the Grassmannian ones in the
target space.  The same symmetries have an image also at the level of the
base-space $(t,\, \theta ,\, {\bar\theta})$. In QM, where
$\theta,{\bar\theta} = 0$, all these symmetries are lost. Now let
us ask ourselves if there is a universal symmetry of CM that is always
lost in QM. The answer is  yes !, \cite{ennio84}, and it corresponds to the
{\it universal} symmetry of rescaling the over-all action. 
That symmetry can be put in the
form of a transformations relating the $\theta$ with the t, \cite{go2000}. 

All this indicates
that some geometrical structure, even if Grassmannian, is behind
the quantization problem.

\section{Acknowledgments}

A.~A. is glad to express his gratitude to the Department of Theoretical
Physics of the University of Trieste where a part of the work was done.
E.~G. would like to thank V.~De~Alfaro, for inviting him to deliver this
talk, M.~Berry, E.~Deotto, D.~Mauro,  M.~Reuter for helpful
discussions and F.~Legovini for much needed help with \LaTeX.

This work was supported by grants from MURST, NATO and INFN.

%%%%%%%%%%%%%%%%%%%%%%%%%%%%%%%%%%%%%%%%%%%%%%%%%%%%%%%
%%%%%%%%%%%%%%%%%%%%%%%%%

\end{document}